# Efficient large-scale STEM-EELS simulations with torched-TACAW

Martin Osmera [a],*, João Vaz [a], Paul M. Zeiger [a,b], Ján Rusz [a]

[a] *Department of Physics and Astronomy, Uppsala University, Box 530, 75121, Uppsala, Sweden*

[b] *Materials Science & Engineering Department, University of Washington United States, 302 Roberts Hall, Box 352120, Seattle, WA 98195-2120, United States of America*

ABSTRACT

The time auto-correlation of auxiliary wave functions (TACAW) method enables efficient simulations of ultra-low-loss electron energy loss spectra (EELS) arising from vibrational and magnon excitations. In practical applications to realistic materials systems, however, TACAW calculations become challenging due to the large system sizes required for models containing defects, interfaces, impurities, or grain boundaries, as well as the substantial computational cost and data throughput associated with molecular dynamics and multislice calculations. Here we discuss a practical methodology for large-scale TACAW simulations and present `torched-TACAW`, a freely available implementation of the TACAW part of the described workflow for efficient STEM-EELS simulations. The overall approach combines molecular dynamics based on foundational machine-learned interatomic potentials, partitioning of elongated supercells, and on-the-fly processing of multislice outputs in order to enable near *ab initio* quality simulations with tractable memory use and data flow. Using rutile $TiO_2$ as a model system, we analyze important numerical aspects of the method, including windowing and supercell partitioning, and demonstrate atomic-resolution STEM-EELS simulations for thick samples.

## 1. Introduction

Since 2014, when the electron beam in a scanning transmission electron microscope (STEM) was monochromated to a full-width at half-maximum of less than 10 meV while maintaining all the advantages of a modern aberration-corrected STEM [1], microscopists started exploring atomic vibrations at high spatial resolution, studying isotopes [2–4], doing local measurements of temperature [5–7], exploring single-atom impurities, stacking faults or grain boundaries [8–10] and reaching the atomic scale [11,12]. Recently electron energy loss spectroscopy (EELS) of phonons with off-axis detectors enabled to explore phonon polarization vectors [13–15], having implications in superconductivity [16]. The monochromated EELS has also enabled to detect another quasiparticle residing in a similar energy range, magnons, which are excitations of the spin sub-system [17].

Electrons interact strongly with materials, leading to multiple scattering effects, which often complicate the direct interpretation of measurements. For that reason simulations are often required to accompany experiments. Simulations of phonon and magnon EELS have a relatively short history, since these scattering events were previously just a part of the flanks of an intense zero loss peak due to insufficient energy resolution. Early simulations were performed in the first-order Born approximation [18–20], which often captured the main spectral features, nevertheless they did not contain electron beam propagation effects. Dynamical diffraction effects were incorporated in inelastic multislice methods [21] as well as Bloch-waves methods [22]. Both of these approaches require knowledge of the individual phonon modes, which becomes computationally demanding for large structure models.

Recently, two alternative simulation methods have been developed. The Frequency-resolved frozen phonon multislice (FRFPMS) method [23] introduced the use of molecular dynamics (MD) simulations for the calculation of structure snapshots containing vibrational modes within a selectable frequency range. Notably, FRFPMS side-stepped the need to explicitly calculate phonon modes, which enabled it to treat large structure models containing tens of thousands of atoms, as is typically needed to describe systems with non-trivial nano-structures like interfaces, systems with defects and similar [24]. In 2025 another approach has been described, named time-autocorrelation of auxiliary wave functions (TACAW; [25]). The TACAW method also utilizes MD and multislice simulations, but uses the elapsed simulation time to derive an autocorrelation function of the beam electron wave function, whose Fourier Transform is proportional to the EELS. In principle, TACAW approximates EELS including multiple elastic and inelastic scattering events, while maintaining or even surpassing the computational efficiency of FRFPMS for large structure models. In addition,

* Corresponding author.
*E-mail addresses:* martin.osmera@physics.uu.se (M. Osmera), joao-pedro.joaquim-vaz.2533@student.uu.se (J. Vaz), paul.zeiger@physics.uu.se (P.M. Zeiger), jan.rusz@physics.uu.se (J. Rusz).

TACAW offers an appealing physical transparency and a relative ease of implementation (see, e.g., a recent work [26]).

In this paper we focus on the TACAW method, particularly from a perspective of performing tractable simulations of realistic large structure models at near *ab initio* accuracy. Under the hood, one needs to obtain MD trajectories of sufficient length and time resolution to carry out such calculations, and then perform a large number of multislice (MS) electron beam propagations to obtain the auxiliary wave functions [27,28] denoted here $\phi_{\mathrm{MD}}(\mathbf{q},t)$, where time $t$ indicates time within the MD simulation. These auxiliary wave functions are then used to obtain $\mathbf{q}$ and energy-resolved scattering intensities given by

$$I(\mathbf{q},E) \propto \frac{\beta E}{1-\exp(-\beta E)} |\phi_{\mathrm{MD}}(\mathbf{q},E)|^2. \quad (1)$$

This is the central equation of the TACAW method [25]. There, $\phi_{\mathrm{MD}}(\mathbf{q}, E)$ is a time-to-energy Fourier transform of $\phi_{\mathrm{MD}}(\mathbf{q},t)$ and $\beta = 1/(k_B T)$ with temperature $T$. The evaluation of intensities as a post-processing of auxiliary wave functions is typically only a small percentage of the total computing costs. Below we outline the main practical challenges associated with these components of a TACAW calculation.

Molecular dynamics simulations leverage the three orders of magnitude of difference between nuclear and electronic masses, treating the nuclei as classical objects, subject to interatomic forces. These forces can be calculated at various levels of sophistication, ranging from simple Lennard-Jones model [29], via empirical potentials [30], harmonic pair-potentials based on force fields calculated by first principles calculations [31], machine-learning models [32], to fully *ab initio* calculated interatomic forces [33]. When the target size of models is in the range of tens to hundreds of thousands of atoms, with current hardware the *ab initio* force calculations are far too demanding. Simple empirical potentials, on the other hand, rarely have enough of degrees of freedom to faithfully describe the vibrational models in all of their complexity. Recently, machine learning models have been introduced, which reach an accuracy comparable to *ab initio* calculations, but at a significantly lower computational costs, making them a natural candidate as the material-specific ingredient of TACAW. The material-specificity of interatomic potentials may pose a challenge as well, since training the interatomic potential is a time-consuming task, typically requiring specialist knowledge.

A convenient solution is offered by foundational models — so called *universal* interatomic potentials, trained on a huge dataset of *ab initio* simulations [34,35]. Several such models have been introduced in recent years, such as M3GNET, ORB, MACE [36–40] and a number of others [41]. Convenience of universal interatomic potentials comes with its own challenge, which is associated with their computational requirements. While they are still significantly faster than *ab initio* calculations, for systems of tens of thousands of atoms they do require substantial memory and computing time. In particular, ORB potentials present an excellent balance of accuracy, speed and memory demands [37,39]. Yet, on a current high-end GPU hardware – an AMD MI250x GPU equipped with 64 GB dedicated high-bandwidth memory (supercomputer Dardel at the PDC Center for High Performance Computing, Stockholm, Sweden) – one can only treat systems with up to approximately 25000 atoms. This is often insufficient for thicker samples. Consider a model 30 nm thick with lateral dimensions of ca 4.5 nm × 4.5 nm, which is typically needed in order to prevent beam spreading to neighboring periodic copies of the supercell. Such a supercell would, for example, contain approximately $10 \times 10 \times 100$ unit cells of a rutile $TiO_2$, in total hosting 60 000 atoms, i.e., substantially more than is possible with above-mentioned GPU. Structure models with defects or multilayers can easily contain even more atoms. Running such calculations on a CPU is certainly possible, but would require a runtime of many days, unless running in parallel on a supercomputer. Therefore if near *ab initio* level of accuracy is desired, a refinement of the current approach is needed.

Another challenge is related with the sheer amount of data flow in a STEM-EELS TACAW calculation. Consider scanning an area of 1 nm$^2$ with a scan step of 0.2 Å in both directions. This calls for 50 × 50 beam positions. For each of the beam positions, in order to obtain spectra with good statistical accuracy one needs on the scale of 1000 structure snapshots from a MD trajectory. A typical multislice grid spacing of 0.05 Å for an above-mentioned supercell will sample the electron beam wave function on a grid of 900 × 900 pixels. Storing double precision complex numbers that cost 16 bytes, this scenario leads to 2.5 million auxiliary wave functions (coming from 2.5 million multislice electron beam propagations) with a total volume of 29 TB. Writing this amount of data for post-processing is not only very time consuming, but with today's hardware it is typically well out of reach for personal computers. A clever organization of the calculation steps, allowing on-the-fly processing and aggregating of data is required. Such calculations are however perfectly suitable for parallel computing environments and GPUs, since fast Fourier transforms (the core of the multislice computations) are performed extremely efficiently on GPUs and individual multislice runs are largely independent of each other.

In the following sections we describe our approach to the challenges outlined above, implemented in the free open-source software `torched-TACAW` [42]. Section 2 presents the practical workflow, including molecular dynamics, supercell partitioning, and the multislice-TACAW implementation. Section 3 then examines the main numerical choices and approximations, such as windowing and partitioning, and validates them using simulations of rutile $TiO_2$. Finally, Section 4 demonstrates atomic-resolution STEM-EELS simulations for a thick $TiO_2$ sample.

## 2. Methodology

In this section we describe the practical steps required to carry out TACAW simulations for vibrational STEM-EELS in large structure models. We first discuss the generation of molecular dynamics trajectories, including the choice of interatomic force models and a partitioning strategy that makes large simulations tractable on currently available hardware. We then present `torched-TACAW`, our implementation of the multislice-TACAW workflow, with emphasis on the data flow and algorithmic choices needed to keep the calculations computationally efficient. Readers primarily interested in the physical conclusions and numerical validation may wish to skim the more implementation-oriented parts of Section 2.2 and proceed directly to Section 3.

### 2.1. Molecular dynamics

Within a practical realization of TACAW, a time-dependent electron exit wave function is approximated by auxiliary wave functions for structure snapshots corresponding to given time points. These structure snapshots are obtained from a MD trajectory. In MD simulations, accurate interatomic forces are typically obtained either directly from *ab initio* calculations or from machine-learned interatomic potentials (MLIPs) trained on such data. MLIPs are the recommended method of choice, given their near *ab initio* accuracy and orders of magnitude better computational performance than *ab initio* MD for large systems. Whenever available, dedicated MLIPs are typically the most accurate and also computationally efficient approach. Nevertheless, dedicated MLIPs are rarely available for new systems, especially for systems with non-trivial nano-structure. Training MLIPs is a non-trivial and time-consuming task that typically requires specialist expertise. In such cases, so-called foundational models provide a general solution. Being trained on many millions of *ab initio* calculations of systems containing elements across the whole periodic table, they are capable to predict interatomic forces for arbitrary structure models. In our workflow we adopted the ORB potential [37], similar to Ref. [26].

Before we proceed further, we need to mention certain aspects of MD calculations with non-conservative force fields. In these cases,

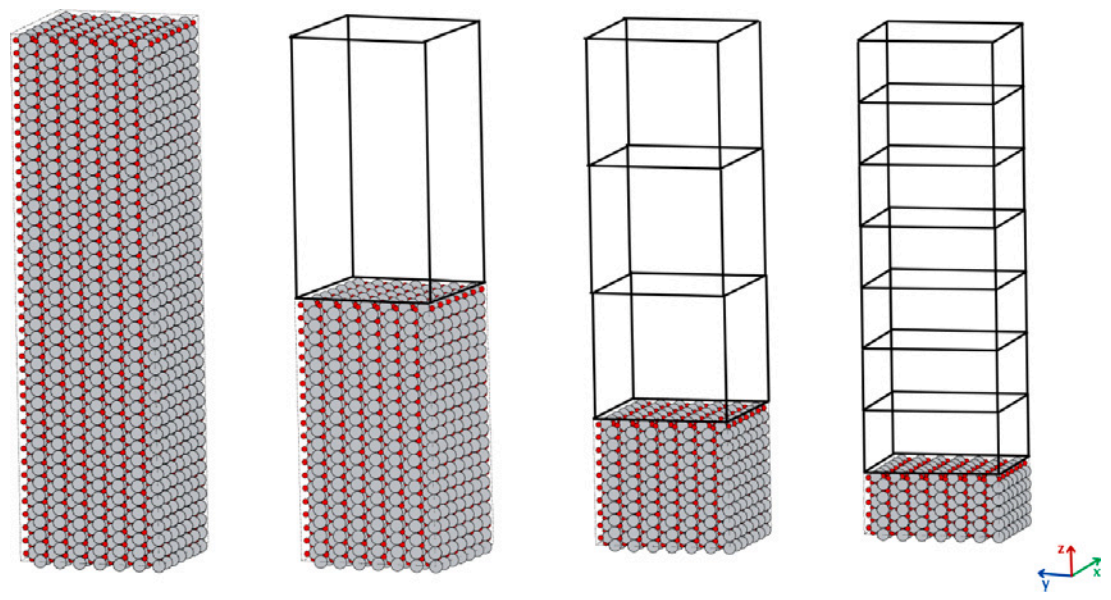

**Fig. 1.** Visualization of a $TiO_2$ supercell (left) and its three different partitionings: in halves, quarters and eights.

the use of a thermostat is necessary in order to counterbalance temperature drifts [43]. We have considered the Langevin, Nosé–Hoover chains (NHC) and stochastic velocity rescaling (SVR), also known as Bussi–Donadio–Parinello, thermostats. Sufficiently strong coupling is needed to maintain a temperature within a few percent of the target temperature. In $TiO_2$ systems explored here, the time constant of 100 fs in each of the three thermostats allowed to maintain an average temperature within 2%–3% of the target temperature of 300 K. We have observed rather strong blurring of the spectral features with the Langevin thermostat, which is a well-known consequence of local thermostats which act differently on each atom, thereby disturbing the system dynamics. Global thermostats, such as NHC and SVR, act on the kinetic energy of the whole system and consequently provide well-defined sharp spectral features. For this reason, thermostats like NHC or SVR are more suitable for calculations of dynamical properties [44], including TACAW. In addition, the stochastic nature of SVR provides an advantage over the deterministic NHC, which can have problems with non-ergodicity [45,46].

It should be kept in mind that a TACAW calculation reproduces only the physics contained in its ingredients. Foundational potentials of the kind used here are short-ranged, and therefore do not describe the long-range electrostatics associated with polarization and charge transfer, which may be significant in polar materials and particularly at defects and interfaces. Potentials capable of treating these effects are under active development [47], and since `torched-TACAW` (Section 2.2) takes the atomic trajectory as its input, it can make direct use of them as they mature. Polarization likewise acts on the scattering itself, through the electric fields experienced by the probe electron; the neutral-atom projected potentials used here do not include this contribution, which could be added either at the level of the potential construction or by substituting a multislice engine that accounts for it, `torched-TACAW` not being tied to a particular engine. More generally, every aspect of the simulation, including the size of the supercell, must be chosen so as not to distort the physics of interest; for defect-related modes, for instance, the supercell must be large enough that the defect does not interact with its periodic images.

Further, we note that we have observed strong numerical sensitivity of spectral intensities in particular voxels of the simulated vibrational EELS data $I(\mathbf{q}, E)$ when using ORB.v2 force fields including D3 corrections [37,48]. The exact reason is not known and is left for future study. In the rest of this work we proceed with force fields without D3 corrections.

ORB is implemented on top of the `PyTorch` library, which allows to run inference either on CPUs or a GPU. The latter provides a significant speed boost, but it may run into problems with insufficient memory, particularly in typical desktop settings and consumer GPUs.

*Partitioning of the supercell*

One solution to this problem lies in partitioning of the structure model. Except for two-dimensional materials, a typical supercell has a "noodle" shape, i.e., being few tens of nanometers long along the beam propagation direction, while the two lateral dimensions are on the order of few nanometers. Since inelastic phonon scattering is primarily due to processes with momentum transfers perpendicular to the beam direction, mildly reduced accuracy in the phonon dispersions along $q_z$ is unlikely to strongly influence inelastic phonon scattering cross-sections. For this reason, we propose to partition the structure model along the $z$-direction into segments of lower effective thickness, stacked together, as depicted in Fig. 1.

We can reduce the size of the supercell that needs to be simulated by MD in this way. Assuming that the system is a homogeneous bulk material, as in the $TiO_2$ example above, we can simulate the above-mentioned $10 \times 10 \times 100$ supercell of $TiO_2$ as four independent $10 \times 10 \times 25$ MD trajectories, stacked on top of each other. Each of the smaller sub-supercells would contain 15000 atoms, which is manageable with the available hardware. Each of the sub-supercells would host correlated atomic displacements as dictated by the interatomic potential, but there will be a discontinuity in the pattern of excited phonon modes in the three planes where the sub-supercells are connected. Section 3 shows that the error introduced by this approach is negligible in typical scenarios.

This partitioning procedure is in a certain sense a generalization of the approach commonly used in Einstein model multislice calculations of thermal diffuse scattering, where multiple variants of a slice potential are pre-calculated, each having a different random pattern of atomic displacements [49,50]. Then, the whole supercell potential is assembled from these variants of slices, picked at random. Here, supercells with correlated motion could be re-assembled from sub-supercells picked at random from smaller MD simulations. This would likely be an efficient way of calculating thermal diffuse scattering within the frozen phonon or quantum excitations of phonons approach with correlated atomic motion. We leave this idea as a suggestion for future work and bring our focus back to TACAW, which requires time-correlations of atomic positions in densely sampled trajectories of sub-supercells.

Finally, a technical note: within our example, one could perform several independent MD simulations with smaller structure models and stack them together. As an alternative, one may perform a single longer MD trajectory and, for each averaging *chunk* in the TACAW calculation (see Section 2.2), construct the stacked supercell from randomly selected segments of that trajectory. Here we will present results for the first strategy, since the hardware that we use contains nodes with 8 independent GPUs, so it is convenient to run up to 8 independent MD calculations in parallel (or more, when using several computing nodes).

For the next subsection, we assume that we have managed to obtain an MD trajectory of a given size, which is sufficiently long for statistically significant TACAW simulations.

### 2.2. torched-TACAW

After a suitable trajectory has been generated, the remaining task is to evaluate the TACAW signal by multislice propagation through the time sequence of atomic configurations; see Eq. (1). In this subsection we present `torched-TACAW`, our implementation for resource-efficient large-scale TACAW STEM-EELS simulations. We first outline the overall architecture and design choices of the code, and then summarize the key implementation details that control performance, memory usage, and data handling. Readers who are mainly interested in the methodological ideas rather than the software design may choose to read only the overview in Section 2.2 and skip the more technical parts of the implementation details below.

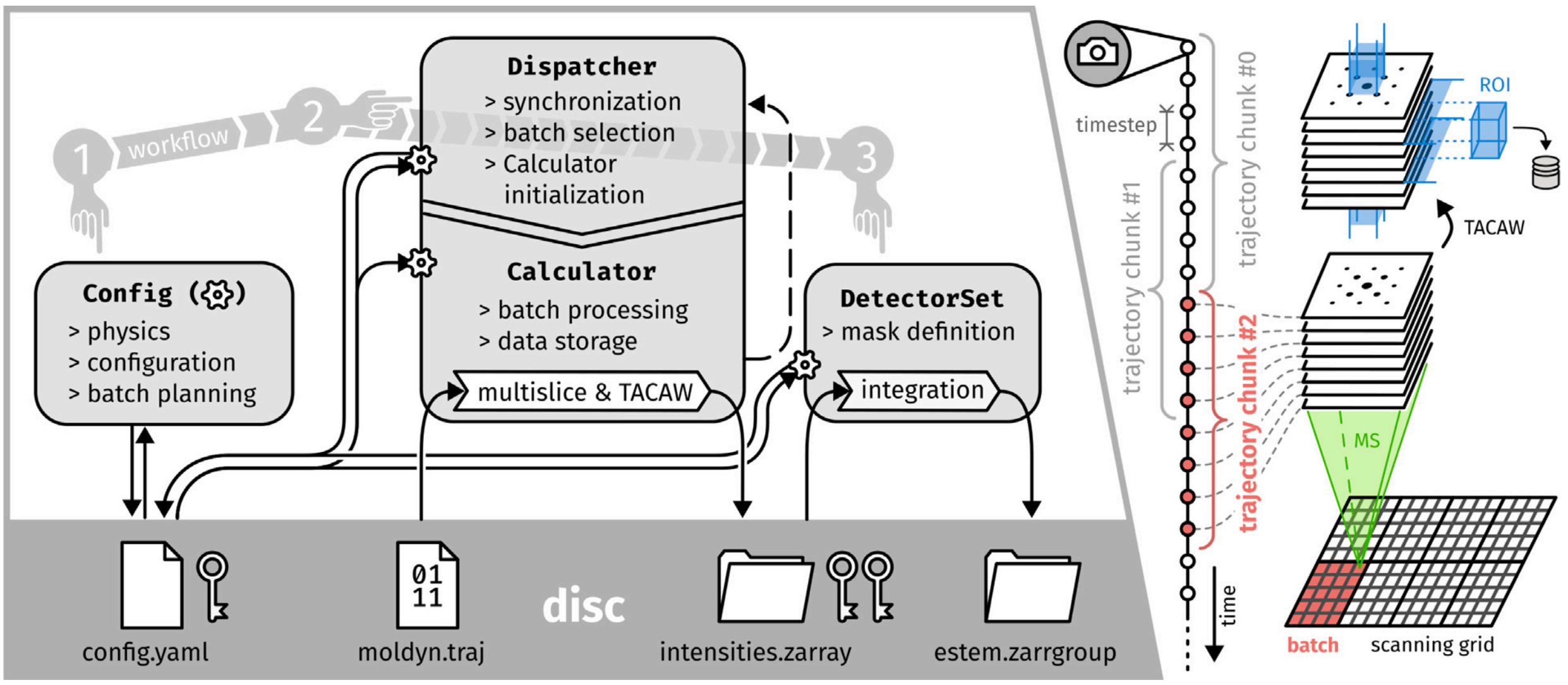

**Fig. 2.** Overview of torched-TACAW's architecture. Left: Core objects and their interactions and dataflow. Typical user's workflow consists of consecutive use of three objects in gray-shaded panels. Right: scheme of one-batch computation and visualization of the final 5-dimensional (pre-detection) dataset. In the intensities.zarray dataset, for each point in scanning grid, data from the region of interest (ROI, blue cuboid) in time and reciprocal space are saved into intensities.zarray.

#### 2.2.1. Architecture & design choices

As the multislice algorithm consists inherently of repeated large-matrix multiplications and Fourier transforms, it is natural to expect computational advantage from GPUs compared to CPUs [50]. Easier availability of GPU-centric nodes on HPCs in recent years motivated us to try to exploit this for the benefit of our calculations. Torched-TACAW was designed to be computationally efficient, transparent, and easily configurable in Python-based workflows. It is not intended as a general-purpose (S)TEM simulation package — comprehensive tools of that kind, such as abTEM [51], are already available — but as a dedicated engine for carrying out TACAW calculations at scale.

We employed Hamish Brown's well-developed py_multislice (pyms) [50] as the multislice engine. The pyms internally uses the PyTorch [52] library for GPU acceleration. PyTorch is, however, to a high extent platform-agnostic and can be set to work on CPUs too. Pyms can conveniently exchange structure data with the ASE library [53]. For data storage the format Zarr v3 was chosen [54].[1]

The projected potentials are constructed within the independent-atom model following the FDES algorithm [55]: atoms are assigned to slices according to their coordinate along the beam direction and, within each slice, rasterised onto the lateral grid and convolved with the electron scattering factor of the corresponding element, taken from the parameterization of Lobato and Van Dyck [56]. Lateral displacements are thus resolved with sub-pixel accuracy, whereas all atoms of a slice are projected into its entrance plane, so that motion along the beam is effectively disregarded. We accept this deliberately, in exchange for a small number of slices, since vibrational EELS is only weakly sensitive to such motion. The slice boundaries are then chosen so that each atomic layer lies in the middle of its slice — for the rutile structures used here, ca 0.74 Å from the nearest boundary, well beyond the root-mean-square thermal displacements of about 0.1 Å. Atoms thus never move between slices, which would otherwise introduce a spurious time dependence and corresponding artifacts in the spectra.

As the problem is embarrassingly parallel, it is possible to divide the large scale computation into computational batches that can be processed simultaneously. Torched-TACAW can be run on an arbitrary number of GPUs in parallel without any need for direct GPU–GPU communication (synchronization is achieved plainly by file-locking on disk).

As was pointed out above, one of the crucial factors is large data traffic. Torched-TACAW keeps most of the data throughput in (V)RAM ((Video) Random-Access Memory) and communicates with the hard-disk only in order to load configuration files for the initialization of the Calculator object (see below) in the beginning of each chunk processing and at its end to save the final result. This makes the calculations faster and removes the need for intermediate storage of TBs of data in wave functions that needed to be postprocessed in the older workflows [17,25].

The code of torched-TACAW is organized into four main objects: Config, Calculator, Dispatcher, and DetectorSet. See Fig. 2 for the general overview of the code architecture, workflow and data-flow. More details of their working are discussed in Section 2.2.2.

Config is responsible for keeping track of physics and computation organization. A large part of the code's performance is hidden in the partitioning of the whole calculation into computational batches that can be processed in parallel. For optimal use of available resources, it is necessary to choose the partitioning properly.

Calculator is responsible for performing the computation for each batch, loading trajectory data from the disk and dumping the corresponding part of the final result to disk.

Dispatcher is responsible for the selection of a batch to be computed, keeping track of its progress and running the Calculator. Several Dispatcher objects can run in parallel. Between each other they are synchronized by locking the config.yaml file during the critical moments of operation. By running more Dispatcher objects in parallel the user can speed up the calculation several times.

The result of the whole calculation, after all batches were successfully computed, is stored in a zarr array on the disk. In most situations, we are interested in the signal integrated from a region corresponding to some specific detector aperture. For this purpose, the object DetectorSet is implemented. When given description of all desired detector geometries, it computes the corresponding energy-filtered STEM images. Currently three types of detectors are implemented: *circular* and *annular* — with arbitrary center and radii — and *custom*, which can be an arbitrary detector provided by the user either as numpy array directly or as a path to a file containing the corresponding array. All three perform a complete reduction over reciprocal space, yielding a single intensity per scan position and energy loss. Detection schemes that retain momentum resolution along one direction, such as those commonly using a slit aperture, can amount to only a partial reduction and are not expressible in this form; the corresponding energy–momentum maps, e.g. as in Figs. 3 or 4, are instead obtained by manual post-processing the stored $I(\mathbf{q}, E)$ dataset directly.

[1] Note the incompatibility with zarr v2.

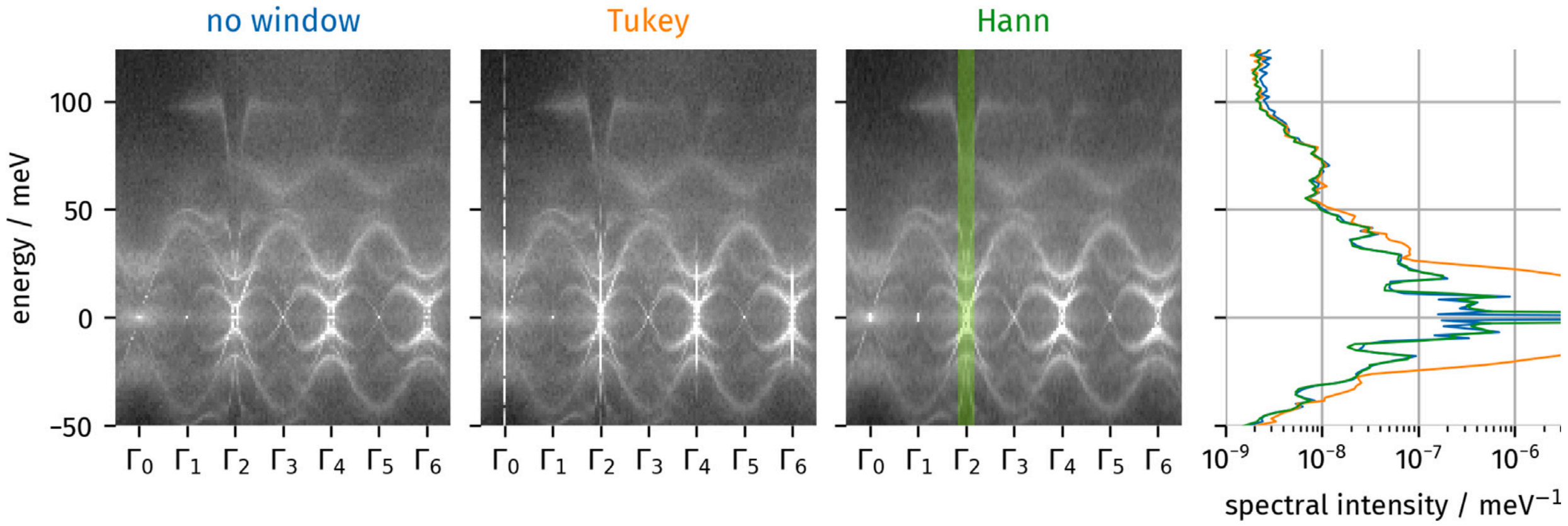

**Fig. 3.** Section of the $I(\mathbf{q}, E)$ datasets for 3 nm thick structure model of $TiO_2$ with lateral dimensions of 9.3 nm × 9.3 nm, calculating using no window function (left), Tukey window with $\alpha = 0.1$ (center) and Hann window (right). The line plot on the right side shows spectra from the highlighted region calculated with the three window functions, respectively.

The energy-resolved diffraction patterns stored in the zarr array represent the scattering probability; the only noise present is statistical, arising from the finite length of the MD trajectory. Shot noise is not included by default, since it depends on the electron dose rather than on the specimen or the beam optics. It can be added for a specified dose either by the `DetectorSet` object, which applies it to the diffraction patterns before integration, or via the function `add_shot_noise()` in the `tools` submodule when post-processing $I(\mathbf{q}, E)$ directly.

#### 2.2.2. Details of implementation

*Configuration.* `Config` object is basically a nested dictionary (directly accessible by `Config.config`) wrapped into an object with various methods generating the partitioning scheme; keeping track of physics, including units and scaling; and simulation grids and coordinates. It is also responsible for keeping track of the location of data on the disk. `Config` has its image stored in the configuration file, with the default filename `config.yaml`. It is a human readable file in the yaml format [57] directly corresponding to the inner `Config.config` dictionary. `config.yaml` is also the reference for keeping track of progress (i.e. keeping track of which computations batches were not yet started, are being processed, or have been successfully finished) and as such can be directly observed by the user. The `Config` object can be instantiated either by direct initialization from keyword arguments (when setting up the calculation) or by loading the `config.yaml`. The user can conveniently use the `Config` object for direct coordinate retrieval, e.g., for postprocessing or plotting of results.

*Processing of a batch.* The `Calculator` object keeps its own instance of `Config` in `Calculator.config`. It needs to be initialized with the specific `batch_id` of the batch to be computed. By design, the user does not need to interact with the calculator directly, as it is the job of the `Dispatcher` to instantiate the `Calculator` object and order it to perform the computation. The batch is characterized by a subset of scanning points and an interval from trajectory (stored on disk in ASE `.traj` binary format). In each batch the `Calculator` performs the multislice calculation on each snapshot in the batch's trajectory interval. It collects the time-consecutive exit wave functions from the defined region of interest (ROI) of reciprocal space and afterwards performs fast Fourier transforms (FFT) of the wave functions from the time to the frequency (or energy) domain. Then the squared modulus of the resulting energy-dependent auxiliary wave function is computed and multiplied by the quantum statistics correction factor, see Eq. (1). The result of this process is the energy-resolved diffraction pattern $I_i^\sigma(\mathbf{q}, E)$ corresponding to the $i$th trajectory interval (chunk) and scanning (sub)-region $\sigma$ — with both $i$ and $\sigma$ defined by the `batch_id`. Given total $N$ different trajectory intervals to be processed, $\frac{1}{N} I_i^\sigma(\mathbf{q}, E)$ is then added to the corresponding part of the zarr-array on the disk. The resulting array stored on disk is thus an average over all trajectory chunks and the union of all scanning subsets

$$\bigcup_{\sigma \in \Xi} \sum_{i=1}^{N} \frac{1}{N} I_i^\sigma(\mathbf{q}, E), \tag{2}$$

where $\Xi$ is the set of all scanning batches (see Fig. 2).

*Batch dispatching.* The `Dispatcher` object keeps its own instance of `Config` in `Dispatcher.config`. From there it dispatches a not yet started batch for computation, changes its status to "in-progress", and instantiates and runs the `Calculator` object for a given batch. After successful completion of the calculation it updates the status of the batch in `config` to "finished". When one batch is finished, `Dispatcher` attempts to start a new not yet dispatched batch.

*Division of the whole computation into batches.* As discussed above, the typical total data volume that needs to flow through the computation is often far too large to perform the whole calculation in one run within (V)RAM. The main power of `torched-TACAW` is in automated division of the whole computation scheme into computational batches. One computational batch consists of one trajectory-interval (chunk) and a subset of points in scanning grid (scanning batch) — see the right side of Fig. 2, where one computational batch is shaded red. This subset's shape in the current implementation needs to divide the shape of the whole scanning grid without remainder.

*Region of interest (ROI) slicing in reciprocal space and energy.* To have a well-converged diffraction image at sufficient angular resolution, the multislice grids need to be relatively large. Most of reciprocal space, however, is not interesting for the analysis of the results as we are often interested only in region with a radius of $\sim 100\,\mathrm{mrad}$ around the central beam image. In order to save computational resources, the `Calculator` retains only a small ROI from the whole reciprocal space and thus also reduces the RAM usage often by one or two orders of magnitude. This reduction of the computational domain enables a large improvement in the code's performance.

Similarly, one may only be interested in a certain interval of energy losses, which defines a ROI in energy. In the interest of further reducing the memory footprint and data retention requirements of the computation, only the ROI in energy is retained.

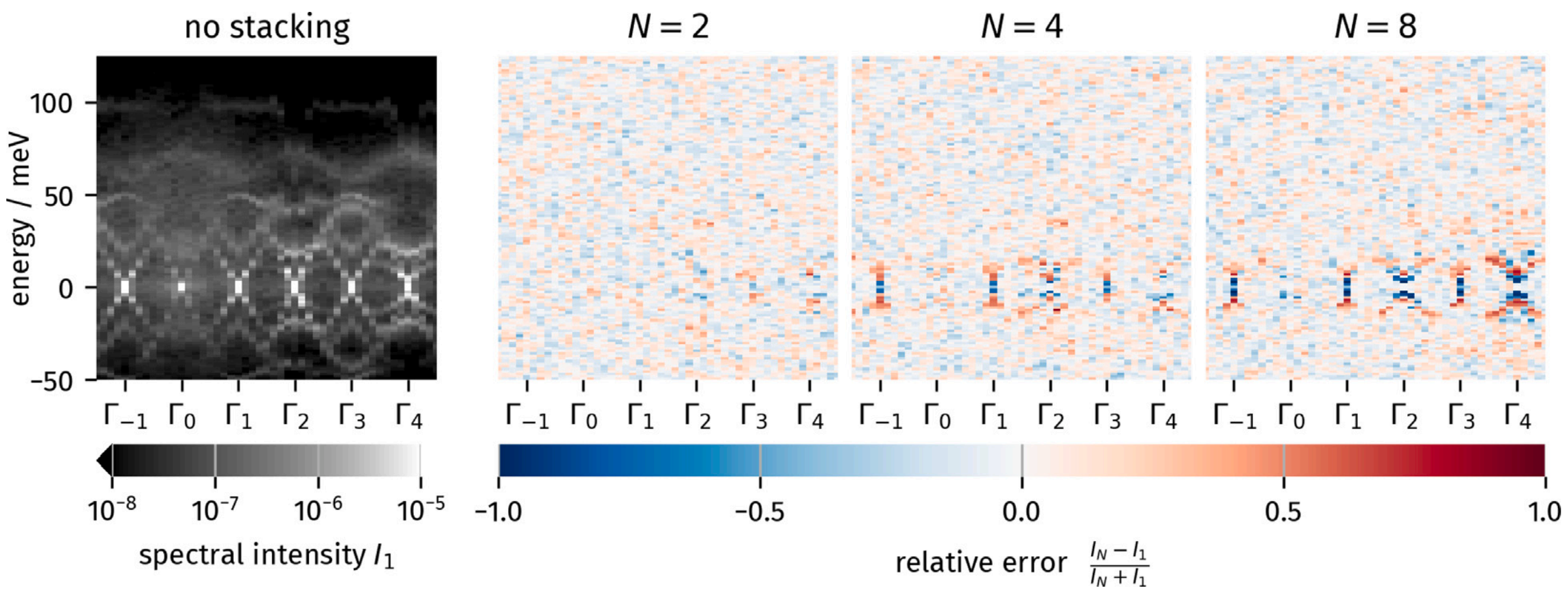

**Fig. 4.** Demonstration of errors introduced by supercell partitioning. The left panel shows a scattering angle-resolved spectrum of full supercell trajectory. The right panel shows the pixel-wise relative differences introduced by partitioning the supercell into $N = 2, 4$, or 8 subcells and running MD separately on each of them. Most of the error introduced is noise-like.

*Welch method: trajectory chunk overlap & windowing.* A direct TACAW evaluation on the whole MD trajectory would generally be impractical. First, storing and processing time-correlated auxiliary wave functions for the full trajectory would lead to operational memory requirements that are often well beyond the reach of available hardware. Second, if there is a time-varying frequency contents in the trajectory, which is the case in a thermostatted MD simulation, it is more advantageous to divide the trajectory into shorter segments, perform the TACAW analysis on each segment separately, and subsequently average the results. More precisely, we partition the trajectory into partially overlapping segments of length $L$. Since the MD trajectory, as well as its arbitrary sub-segments, are in general aperiodic, we use window functions (see below) to taper the time series and reduce the spectral leakage artifact. This procedure is directly analogous to the classical Welch method of spectral density estimation [58].

In `torched-TACAW`, the overlap is controlled by the parameter `chunk_overlap`, denoted here by $v$. This parameter directly specifies how many times an average snapshot (i.e., excluding those near the beginning and end of the full trajectory) is reused in different chunks. If a chunk indexed by $n$ starts at snapshot $s_n$, then the next chunk starts at $s_{n+1} = s_n + L/v$, using remainder-less division. The default value is $v = 2$. `Torched-TACAW` also allows the user to specify a stride for snapshot reading from the `.traj` file.

After multislice propagation has been performed for all snapshots in a given segment, the resulting time-correlated exit wave functions are multiplied by a window function before the Fourier transform. By default, `torched-TACAW` uses a Hann window, although several other window functions are available. The purpose of the window is to suppress artifacts caused by the finite length of trajectory chunks. Further discussion of the choice of window function is given in Section 3.1. The windowed sequence of auxiliary wave functions is then Fourier-transformed in time, modulus squared, and multiplied by the quantum-statistical correction factor, see Eq. (1).

*Data storage.* The config-file (`config.yaml`) is a yaml-formatted text file [57] containing the image of the `Config` object's inner dictionary. Through this file, the `Dispatcher` objects also synchronize their work and keep track of the overall calculation progress.

The result of a TACAW-multislice calculation — the energy-resolved diffraction pattern for every beam position — from the defined ROI is stored on the disk as a zarr array (`tacaw.zarray`). The order of axes of this array is [dummy, energy, scan-x, scan-y, $k_x$, $k_y$]. The dummy axis is left for future use, e.g., for beam tilting, different property analyses, or time-resolved calculations. The next axis is the energy axis followed by axes of scanning positions (scanning positions are defined by the user and do not necessarily need to follow an orthogonal window or be aligned with the structure model's axes, which need to be orthogonal). The last two axes, $k_x$ and $k_y$, correspond to the momentum transfer in reciprocal space. The coordinates of all axes are easily accessible from the `Config` object by corresponding getter functions.

The results of energy-resolved STEM calculated by `DetectorSet` from the data in `tacaw.zarray` are stored in a zarr group — a set of multiple zarr arrays — (by default `estem.zarrgroup`). The Zarr array for each detector is labeled by the detector's label. Axes of these arrays are [energy, scan-x, scan-y].

*Logging.* Detailed logging is available through the use of Python's logging module. It can be useful for tracking progress of the calculation or for checking for possible errors.

*Normalization.* All data are normalized "per voxel", so that in order to get physical probabilities, one needs to only sum over the values in the corresponding array segments. The physical dimension of voxels can be retrieved from `config`.

*User interface.* `Torched-TACAW` is installable as a python module. `Torched-TACAW` is not equipped with a graphical user interface. All interaction is intended from python scripts or interactive sessions. The user should interact directly with three objects: `Config`, `Dispatcher`, and `DetectorSet`. We present sample code snippets in Appendix B.

## 3. Avoiding artifacts in TACAW calculations and algorithmic refinements

In this section we examine the main numerical aspects that influence the quality and reliability of TACAW simulations in practice. In particular, we discuss the role of window functions in suppressing spectral artifacts caused by finite-time Fourier transforms and assess the accuracy of the supercell partitioning strategy introduced above.

### 3.1. Windowing

We illustrate the effect of window functions on the numerical results obtained for a thin sample of $TiO_2$. A 20 × 20 × 10 supercell of $TiO_2$ has been constructed, having dimensions of 9.3 nm × 9.3 nm × 3.0 nm, consisting of 24 000 atoms. The numerical grid in multislice calculations has been set to 1680 × 1680 pixels, corresponding to a real-space grid spacing of about 0.055 Å. The relatively large lateral size offers good angular resolution with **q**-space grid spacing corresponding to 0.52 mrad. We consider a plane-wave electron beam accelerated by 60 kV voltage and propagating parallel to the $c$-axis of the $TiO_2$ cells. Since each unit cell consists of two atomic layers perpendicular to the beam direction, the supercell was split into 20 slices of equal thicknesses in the multislice calculations.

The TACAW calculation was performed using Welch's method with the overlap parameters set to $\nu = 2.0$ and chunk lengths of 200 snapshots with a time-step of 15 fs. These settings lead to an energy sampling of 1.4 meV. The results represent an average over 31 chunks. The molecular dynamics calculation was performed at room temperature using the SVR thermostat [59] with a time-step of 1 fs for a total length of 51 ps, from which the first 3 ps were discarded as a thermalization period. The time constant of the thermostat was set to 100 fs. As was mentioned above, the foundational model ORB.v2 has been used [37] to describe the interatomic forces, without the D3 corrections.

To save computing time and storage, the ROIs for **q**-space was set to a range of ±100 mrad in both $\theta_x$ and $\theta_y$ scattering directions. For the energy range, the ROI was set to ±30 THz, safely covering the range of single-phonon excitations. The batch size for a plane-wave beam is trivially $1 \times 1$ beam positions. These calculations were executed on two compute nodes, each containing 8 GPUs. The total calculation time was approximately 6 min and resulting data size was 191 MB — a zarr array containing an array of $181 \times 383 \times 383$ of double-precision real values.

In the context of atomic vibrations, frequencies of vibrational modes rarely fit on frequency grids in Fourier-transformed chunks, therefore spectral leakage is expected. From a numerical perspective it is caused by the discontinuity of the chunk across its boundaries, since Fourier transforms implicitly assume that the signal is periodic with a period equal to the chunk length. This discontinuity can be counter-acted by window functions, such as Tukey, Hann, Blackman–Harris and others, which multiply the signal by a smooth envelope that tapers to zero at the boundaries of the chunk's time interval. Tukey window has a form of cosine starting from zero and reaching one, continued by a flat segment of amplitude one, followed by another cosine segment smoothly decreasing from one towards zero. The sum of fractional widths of the two cosine segments is a free parameter of the Tukey window, conventionally called $\alpha$. Hann window can then be introduced as a special case of Tukey window without the flat central segment, i.e., with $\alpha = 1$. Using no window is often referred to as a "rectangular window", consisting of 1's in all of the bins of chunk's time interval. Effectively that corresponds to a Tukey window without the cosine segments, i.e. with $\alpha = 0$.

Intuitively, a Tukey window with a low $\alpha$ may appear particularly suitable, since it alters the data in a very limited way. Nevertheless, Fig. 3 demonstrates that this intuition is to some extent misleading. Note the intense vertical stripes appearing at the Gamma points. These are artifacts of the Tukey window ($\alpha = 0.1$), which has a relatively low dynamic range and zeros of its Fourier transform's side-lobes in general do not fall on the frequency bins. This leads to spreading of the intense delta-like elastic scattering peak into other energy bins. For the more distant Bragg spots, their lower intensity in combination with relatively fast decay of the side lobes of the Tukey window limit the vertical stripe artifacts to a narrower range of energies.

Not using window functions (i.e., using a rectangular window) also leads to artifacts, which are nevertheless more subtle in Fig. 3. A weak vertical double-stripe artifact appears at higher energy range. It occurs in the pixels adjacent to the $\Gamma$-points. This additional intensity manifests itself in the high-energy tail of the EELS spectra, where the spectrum calculated without window functions over-estimates the spectral intensity. Despite that this rectangular window has effectively even more intense side lobes than the Tukey window, the zero loss peak inherently falls exactly into the zero frequency bin and in that particular case with rectangular window one avoids the spectral leakage. Nevertheless, in other **q**-points, in general, the spectral leakage artifact will be present. Considering the strong inelastic scattering intensities for long wave-length acoustic phonons and the slow decay of side lobes of a rectangular window, the intensity of higher order side lobes will eventually overshoot the physical background intensity at higher energy losses.

The Hann window leads to spectra free of both of these artifacts, at the cost of mild spectral broadening. The Hann window has lower and faster decaying side lobes than the Tukey window and, being from the family of cosine-sum windows, it also has the property that the zero loss peak does not lead to spectral leakage artifacts. We have compared the spectra calculated with the Hann window to results obtained using other more aggressive window functions, such as the Blackman–Harris, Kaiser, or Dolph–Chebyshev windows. Blackman–Harris window offers an excellent dynamic range at the cost of stronger spectral broadening. The other windows mentioned here, with their parameters set to offer a high dynamic range, show a similar performance. In comparison with the Hann window, with the exception of the zero loss peak, there is little to no visual difference between the results. Side-lobes of the Hann window decay fast enough (proportional to $1/\omega^3$), while phonon EELS decays slower (proportional to $1/\omega$), therefore the high-energy side of the spectra is determined by vibrational intensities, not artifacts. An additional advantage is a lower level of spectral broadening caused by the Hann window, in comparison with the high dynamic range windows mentioned here. For this reason, the default choice of a window function in `torched-TACAW` is the Hann window. Results for a wider range of window functions can be found in Appendix A.

The usage of windows as discussed above reduces the artifacts stemming from the numerical treatment of the modeled system. If the trajectory retrieved from molecular dynamics has unphysical time dependence, this will be naturally reflected in resulting TACAW simulations. In such scenarios, the particular choice of window function can influence how these artifacts manifest. Both suppression and enhancement are possible results. Windowing can thus "cure the symptoms", i.e., the result of TACAW STEM-EELS simulation, but the "illness", in the form of unphysical MD trajectory, can still propagate into the results.

Examples of these artifacts are observable, e.g., in the early simulations performed by the TACAW method, where only rectangular window functions were used. For example in [25] one can easily notice the vertical streaks appearing in $\Gamma$ points of phonon dispersion in article's Fig. 2.

One may ask whether the suppression of spectral leakage affects the relative weight with which bulk-like and localized contributions are represented. The window scales all modes by the same factor, so the ratio of integrated spectral weights is preserved exactly; the accompanying convolution conserves that weight while reducing the peak height of features narrow compared with the main lobe. The side-lobe intensity removed by the suppression is spurious rather than genuine spectral weight, and the price paid is a wider main lobe: for the settings used here it is about 2 meV wide, against about 1.2 meV for a rectangular window, both well below the energy resolution of current monochromated instruments.

### 3.2. Artifacts stemming from unphysical time-dependence in molecular dynamics

It is worthwhile to mention here some practical aspects that one should be careful about, in order to avoid potential artifacts. In general, TACAW uses the time-correlation between snapshots in MD trajectory to retrieve the energy-resolution in STEM-EELS simulation. Any unphysicality in the time evolution of the trajectory will thus appear as artifacts in the retrieved spectra. Here we will in particular mention the effects of center of mass (COM) drift and the problem of "jumping atoms".

Elastic scattering is the dominant scattering mechanism, therefore any even slight time-dependence of the overall crystal position or shape (beyond the small atomic vibrations) can lead to non-negligible energy dependence in the $I(\mathbf{q}, E)$'s elastic channel.

In particular, drifts of the system caused by nonzero momentum of the center of mass, e.g., due to stochastic forces or initialization of velocities by a Maxwell–Boltzmann distribution, lead to additional snapshot-dependent phase ramps in the wave function. After Fourier transformation, this can lead to non-negligible spurious intensities at nonzero energy loss and gain values everywhere in $\mathbf{q}$-space where there is an elastic scattering at $E = 0$. For instance, in plane-wave calculations these artifacts would appear at $\Gamma$ points. This can be prevented by fixing the center of mass of the supercell during the calculation of the MD trajectory.

Take a simplified example of a structure with single species-atoms indexed by $j$ sitting in their respective equilibrium positions $\vec{R}_j$. Without any vibrations, but with all atoms moving in one direction with a time-dependent drift displacement $\vec{d}(t)$, the scattering amplitude in the first Born approximation and assuming the independent atom model can be written as

$$\begin{aligned}\Gamma(\vec{q},t) &= \int \sum_j f\left(\vec{R}_j + \vec{d}(t) - \vec{r}\right) \exp\left(\mathrm{i}\,\vec{q}\cdot\vec{r}\right) \mathrm{d}^3 r \\ &= \sum_j f\left(\vec{q}\right) \exp\left[\mathrm{i}\,\vec{q}\cdot\left(\vec{R}_j + \vec{d}(t) - \vec{r}\right) + \mathrm{i}\,\vec{q}\cdot\vec{r}\right] \\ &= \exp\left(\mathrm{i}\,\vec{q}\cdot\vec{d}(t)\right)\, \Gamma_0(\vec{q},t) \end{aligned} \tag{3}$$

where $f$ and $\Gamma_0$ are the atomic scattering factor and structure factor, respectively, and $\vec{r}$ and $\vec{q}$ are the real and reciprocal space coordinates, respectively. The prefactor $\exp\left(\mathrm{i}\,\vec{q}\cdot\vec{d}(t)\right)$ is the afore-mentioned phase ramp.

Taking the Fourier transform in time to retrieve the energy spectrum then gives

$$\Gamma(\vec{q},\omega) = \int \exp\left(-\mathrm{i}\,\omega\, t\right) \exp\left(\mathrm{i}\,\vec{q}\cdot\vec{d}(t)\right)\, \Gamma_0(\vec{q},t)\, \mathrm{d}t\,. \tag{4}$$

Assuming a constant COM motion $\vec{d}(t) = \vec{v}_{\mathrm{COM}} t$, this leads to a Doppler effect, shifting the frequency in a $\vec{q}$-dependent way. In combination with spectral leakage, this can lead to intensity streaks at nonzero $\vec{q}$.

For a stochastic COM motion, as can be observed for example in Langevin dynamics, the $\vec{d}(t)$ describes a Brownian motion. In consequence, the time-to-frequency Fourier transform will be a convolution of pure $\Gamma_0(\vec{q},\omega)$ with a Fourier transform of the drift-dependent phase factor, which may introduce spurious frequencies into the scattering calculations.

Another case leading to artifacts often occurs when the simulation supercell contains atoms with equilibrium positions with fractional $z$-coordinates that are equal or very close to 0 or 1. During the MD trajectory, such atoms can effectively move through the top or the bottom supercell boundary if periodic boundary conditions are used, i.e., they “jump” across the supercell. In consequence, either such atoms get ignored by the multislice calculation or periodic boundary conditions will fold them back into the supercell, effectively causing atoms to jump through the supercell boundaries from top to bottom or vice versa. Both options induce changes in the scattering intensities that are comparable to or even stronger than thermal diffuse scattering. These can be prevented when constructing the initial supercell. One needs to make sure that there is a sufficiently thick layer of vacuum between the entrance surface and the atom(s) with the lowest $z$-coordinate, as well as between the atom(s) with the highest $z$-coordinate and the exit surface. Considering that the atomic motion in crystals is of the order of 0.1 Å, just a fraction of Ångström is often a sufficient distance. In particular, $TiO_2$ has two layers of atoms, nominally at $z = 0$ and $z = \frac{1}{2}$, with a lattice parameter $c = 2.97$ Å. Shifting the atomic planes to $z = \frac{1}{4}$ and $z = \frac{3}{4}$ and using such unit cell as a basis to create a super-cell, will place the atoms $\frac{c}{4} \approx 0.74$ Å away from both entrance and exit surfaces — a very safe margin to avoid the artifact due to atoms moving across the top or bottom supercell boundaries.

Slow time-dependent changes in the elastic scattering channel can also happen when there are weak forces, such as van der Waals interactions between layers of atoms such as in hexagonal boron nitride. The atomic layers can slowly drift with respect to each other, subject to weak returning forces. Alternatively, if the dimensions of the supercell were not sufficiently relaxed for the target temperature (ideally using NPT calculations), the dynamical matrix in harmonic approximation would reveal a presence of modes with imaginary frequencies. In an MD simulation they may manifest themselves as local distortions in the supercell, which could slowly drift over time. All these effects may be sources of artifacts in TACAW calculations and require a careful consideration, when setting up calculations.

### 3.3. Supercell partitioning

In this section, we demonstrate and validate an approach in which multiple independent trajectories are combined to construct an effective trajectory of a larger supercell, which is then used in a TACAW calculation. To enable direct validation, we select a moderately sized structural model such that the full supercell remains tractable for MD simulations with the ORB potential. This setup allows us to generate a reference TACAW dataset from the full supercell simulation. We then benchmark the errors introduced by partitioning the large system into smaller sub-supercells, running independent simulations, and recombining the resulting trajectories for TACAW simulations. In particular, we construct a supercell of size $8 \times 8 \times 64$ unit cells of $TiO_2$, containing 24 576 atoms. Apart from running MD for this supercell, we will also calculate (1) $N = 2$ independent MD trajectories for $8 \times 8 \times 32$ supercells, (2) $N = 4$ independent MD trajectories for $8 \times 8 \times 16$ supercells and, finally, (3) $N = 8$ independent MD trajectories for $8 \times 8 \times 8$ supercells. From each of these three sets of trajectories we will re-create a trajectory of $8 \times 8 \times 64$ supercell by stacking them on top of each other, see Fig. 1. Importantly, every single trajectory is initialized with random velocities following the Maxwell–Boltzmann distribution and subsequently thermalized; all these trajectories have the same timestep of 1 fs, frequency of storing the snapshots (every 15 fs) and the total number of 50 000 timesteps.

Subsequently, we run TACAW calculations assuming a plane-wave electron beam accelerated by 60 kV, as in the previous section. The lateral size of the supercell is smaller, resulting in a coarser angular resolution of about 1.3 mrad per pixel. Other settings were analogous to the previous section.

The left panel of Fig. 4 shows $I(\mathbf{q}, E)$ dataset from the non-partitioned calculation. Apart from a reduced angular resolution, it shows features matching with Fig. 3, when using the Hann window. The other panels show a pixel-wise relative error of intensities obtained by stacking $N = 2, 4$ or 8 trajectories with model sizes of $8 \times 8 \times 32$, $8 \times 8 \times 16$ and $8 \times 8 \times 8$ unit cells, respectively. Note that `torched-TACAW` has been successfully tested for larger systems with tens of thousands of scanning points.

Most of the observed differences are noise-like without any dominant features. The noise is slightly tinted red, suggesting a small general increase in the background signal. This intuitively corresponds to a slight increase in noise due to the introduction of discontinuities into the supercell. Some structure is observed only in the immediate vicinity of the Bragg spots. The dependence on momentum transfer, i.e., the strengthening of the differences with higher scattering angles, suggests that these differences could stem from an altered representation of long-wavelength acoustic modes with nonzero momentum in the $z$ direction. In Fig. 5 we plotted three spectra from three different circular detectors. Visually, there is no observable impact of stacking on the spectral shapes, with the exception of the lowest energy losses below 10 meV.

In summary, we find that the artifacts introduced by supercell partitioning are minor for most modes and in most scenarios negligible, especially for atomic resolution EELS in an off-axis geometry.

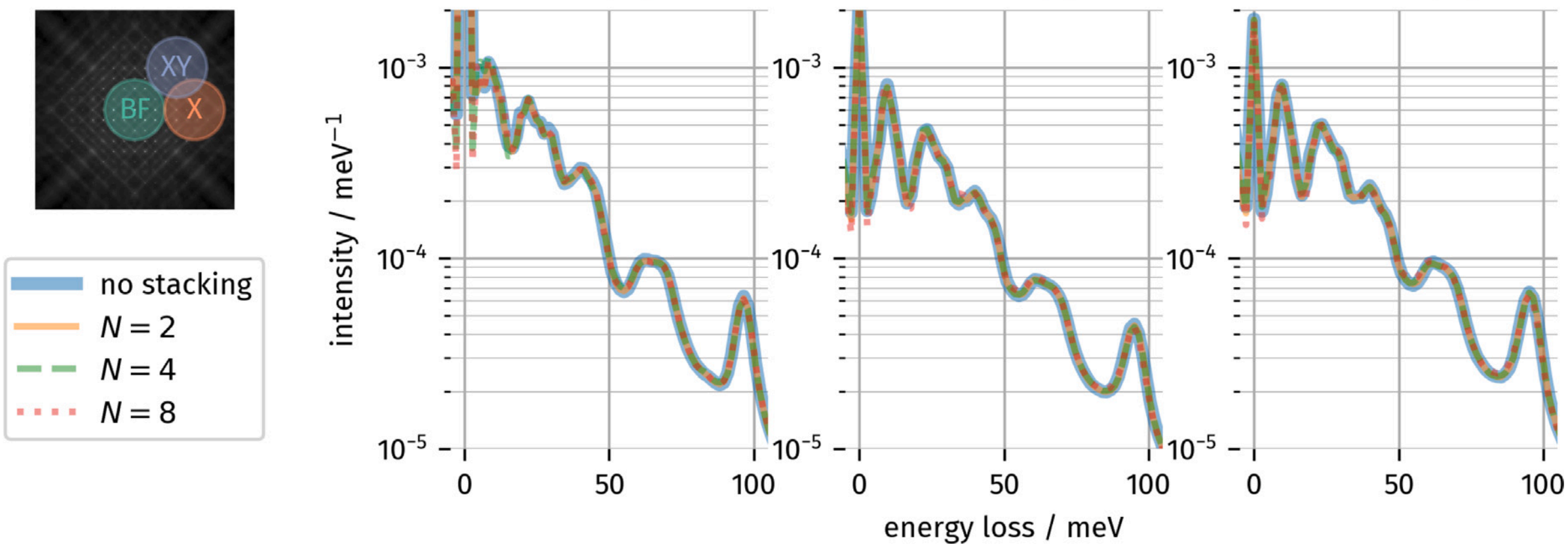

**Fig. 5.** Spectra from three detectors positioned as shown in the upper left diagram. In all three spectra, supercell partitioning has minimal impact. The radii of the circular detectors are 30 mrad and off-axis detectors are displaced 60 mrad from the center.

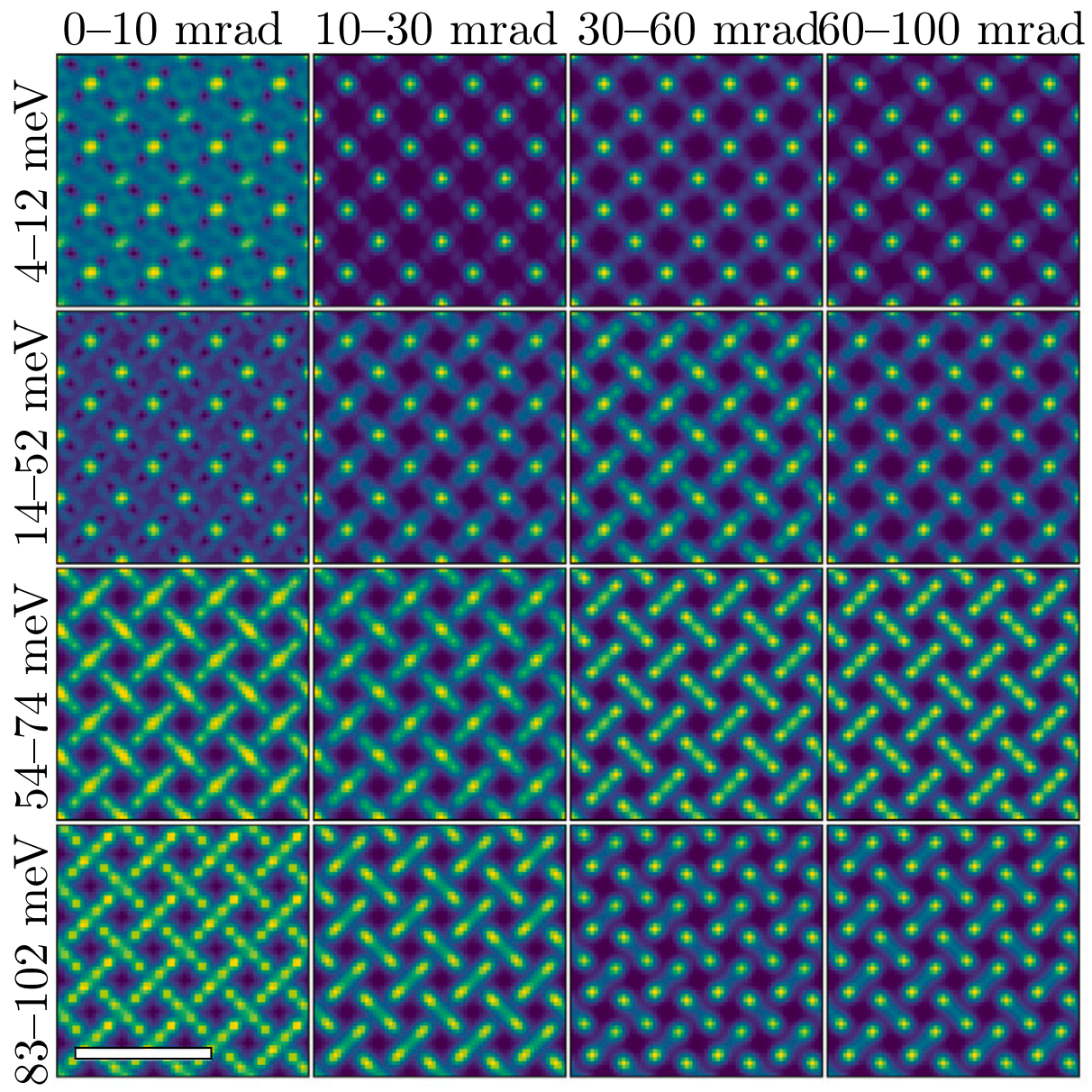

**Fig. 6.** Energy-filtered STEM-EELS images of $TiO_2$ at atomic resolution for a circular bright field detector (left column) and for three annular detectors with indicated inner and outer collection angles. The energy ranges are indicated on the left hand side. The scale bar shows 1 nm.

## 4. Demonstration of STEM-EELS simulation

Having presented the idea of partitioning of the supercell, we proceed to a more challenging case — a STEM-EELS calculation of a 24 nm thick crystal of $TiO_2$ with lateral dimensions of 9.3 nm × 9.3 nm, obtained by stacking 8 supercells of the size that was used in Section 3.1. Effectively, this is a supercell consisting of 20 × 20 × 80 unit cells of $TiO_2$, containing 192 000 atoms in total. All computational parameters and grid spacings from Section 3.1 have been preserved, which means that the model was sliced into 160 slices. In accordance with the ideas laid out in Section 3.3, the snapshots of the 8 sub-supercells originate from 8 independent MD trajectories. New parameters concern the convergence semi-angle of 30 mrad and the scan region sampling an area of one unit cell with 16 × 16 beam positions, corresponding to a scan step of 0.3 Å. This calculation was executed on the same hardware, i.e., using 16 GPU cores in parallel and took 3 h and 40 min. The resulting data was stored in a zarr array of size 48 GB, containing an array of $191 \times 16 \times 16 \times 383 \times 383$ real-valued elements.

In this calculation we have used the `DetectorSet` object, defining 4 different detectors: a bright-field circular detector with a radius of 10 mrad, and three annular detectors with inner–outer angles of 10–30 mrad, 30–60 mrad and 60–100 mrad respectively. Running the detectors on the calculated data results in a zarr store containing four arrays of dimension 191 × 16 × 16, with a total size of only 760 kB. Energy-filtered STEM images for selected energy ranges for these 4 detectors were plotted in Fig. 6. At the lowest energies the phonon EELS is dominated by vibrations of Ti atoms. As the energy range is increased, the visibility of oxygen atomic columns becomes more pronounced and even dominates the image, especially at larger scattering angles — consistent with findings in Ref. [14]. For the bright-field detector we observe a somewhat reduced interpretability of the images due to more pronounced dynamical diffraction effects.

## 5. Conclusions

In this work we addressed the practical requirements for performing TACAW simulations of low-loss STEM-EELS on structurally large models relevant to defects, interfaces, grain boundaries, and other non-trivial nanoscale systems. We presented `torched-TACAW`, an efficient implementation of the TACAW formalism with GPU acceleration. We described our full workflow including the usage of machine-learned interatomic potential, considerations about thermostats, and supercell partitioning, with subsequent employment of `torched-TACAW`. This enables simulations with near *ab initio* quality of atomic dynamics while keeping the computational cost and data flow at a tractable level for modern GPU-based computing.

Using rutile $TiO_2$ as a model system, we demonstrated several key aspects that are important for reliable and efficient calculations. First, the choice of window function has a decisive impact on the suppression of spectral leakage, with the Hann window providing a favorable compromise between artifact suppression and spectral broadening. Second, partitioning elongated supercells into smaller independently propagated sub-supercells introduces only minor, predominantly noise-like errors, while substantially improving the feasibility of simulations for thick samples. Third, the resulting STEM-EELS calculations reproduce the expected energy- and angle-dependent contrast, including the evolution from Ti-dominated low-energy signals towards increasingly pronounced O-column contrast at higher energies and larger collection angles.

These results show how the TACAW method can be implemented for experimentally relevant systems containing up to hundreds of thousands of atoms. We expect `torched-TACAW` to provide a useful platform for the interpretation of monochromated vibrational and, more generally, low-loss inelastic STEM-EELS experiments in complex materials. The development of `torched-TACAW` is ongoing and we are currently working on implementation of new features into the workflow. Notably we are working on seamless incorporation of the thermostated ring-polymer molecular dynamics (TRPMD) into the `torched-TACAW` framework. This is relevant, e.g., for simulations of low-temperature systems allowing to reliably capture quantum effects such as zero-point motion dominating in cryostatic systems [60]. Another feature in development is the inclusion of magnetism-related effects into the multislice inclusive GPU acceleration. When finished it will allow for fast and user-friendly simulations of STEM-EELS experiments on magnetic materials. This will then allow for straightforward treatment of phenomena resulting from spin-lattice coupling [61].

## CRediT authorship contribution statement

**Martin Osmera:** Writing – review & editing, Writing – original draft, Visualization, Validation, Software, Methodology, Investigation, Formal analysis, Data curation, Conceptualization. **João Vaz:** Writing – review & editing, Visualization, Validation, Investigation, Formal analysis. **Paul M. Zeiger:** Writing – review & editing, Conceptualization. **Ján Rusz:** Writing – review & editing, Writing – original draft, Visualization, Validation, Supervision, Resources, Project administration, Methodology, Investigation, Funding acquisition, Formal analysis, Data curation, Conceptualization.

## Code availability

`torched-TACAW` is free and open-source software, available at https://github.com/Martexiwell/torched-tacaw and archived at https://doi.org/10.5281/zenodo.20929266 [42].

## Declaration of generative AI and AI-assisted technologies in the manuscript preparation process

During the preparation of this work the authors used Microsoft Copilot and Claude in order to flag typos and stylistically improve short segments of the manuscript. After using this tool/service, the authors reviewed and edited the content as needed and take full responsibility for the content of the published article.

## Declaration of competing interest

The authors declare that they have no known competing financial interests or personal relationships that could have appeared to influence the work reported in this paper.

## Acknowledgments

We acknowledge the support of the Swedish Research Council (grant nos. 2024-06617 and 2025-04514) and Knut and Alice Wallenbergs' foundation (grant no. 2022.0079). The simulations were enabled by resources provided by the National Academic Infrastructure for Supercomputing in Sweden (NAISS), partially funded by the Swedish Research Council through grant agreement no. 2022-06725, which is also acknowledged for awarding this project access to the LUMI supercomputer, owned by the EuroHPC Joint Undertaking and hosted by CSC (Finland) and the LUMI consortium, where a part of the simulations have been performed.

## Appendix A. More examples for the methodology of windowing

In Fig. A.7 we present a more focused view of artifacts introduced to the simulation by using various windowing functions. Compared to Fig. 3 we showcase a much smaller region of energy and scattering angles. This makes the linear artifacts around the gamma points easier to distinguish from the background. We also feature several alternative windows — the Kaiser window with $\beta = 3\pi$ or $5\pi$, the Dolph–Chebyshev window with attenuation 100 dB, and the Blackman–Harris window. See scipy and its documentation for details about the window functions [62].

Notice how the application of window functions smears noise in the background, compared to the case without a window, i.e., a rectangular window. Otherwise the most prominent feature of all images is the energy-broadening of the $\Gamma$-points. Generally we can conclude that none of these more sophisticated windows delivers a striking benefit. On the contrary, they smear out the fine spectral features more than the Hann window does.

## Appendix B. "How to cook a TACAW" — example scripts & how to set the computational parameters in `torched-TACAW`

In this section, we present sample code snippets for a model system of $TiO_2$ presented in Section 3 above. Assuming `torched-TACAW` is installed in the python environment as `torched_tacaw` we will expect this import

```
from torched_tacaw
    import Config, Dispatcher, DetectorSet
```

to precede any of the following scripts.

### B.1. Setting up the calculation

First step to be done is to setup the calculation. One is recommended to do so by a script similar to the Listing 1.

This script should be run only once. The config is setup by keyword parameters in the `Config` object instantiation. The concrete structure depends on the specific need of the calculation. Users who prefer to work with input files can simply design a universal script that will load input from an input file, e.g., in `yaml` format and unpack the corresponding dictionary into the instantiation call. We refer to the in-code documentation for all necessary details.

By running this script, the config object is created. By including parameter `dump_to_yaml = True` the config is immediately dumped to the config file. This script is designed to be run on the machine, where the actual computation will be performed, i.e., typically a supercomputer. However, one can easily prepare this configuration and file structure locally and then synchronize it to a remote machine.

*Scanning.* `Torched-TACAW` currently offers four different scanning modes: 'unitcell', 'box', 'box_relative', and 'parallelogram'. In all of these, the scanning is performed over a two dimensional grid with grid shape (i.e., the number of points) defined by the user explicitly. Note that the end point is by design "not included" — for example, 5 scan points (as in `scanning_shape`) on interval $[0.0, 1.0)$ would result in points $\{0.0, 0.2, 0.4, 0.6, 0.8\}$. See the documentation for explicit explanation of all possible settings for scanning.

*Trajectory.* The trajectory file must be in a format which ASE can parse through the `Trajectory` object. We suggest using the binary .traj format native to ASE. The trajectory file by default needs to be present while generating the config file, but this can be overruled by the user. More details about this option and other possibilities can be found in the documentation. One needs to pay attention to the simulation time sampling when generating the trajectory file, since it influences the range and sampling of frequencies (energies) in the TACAW simulation.

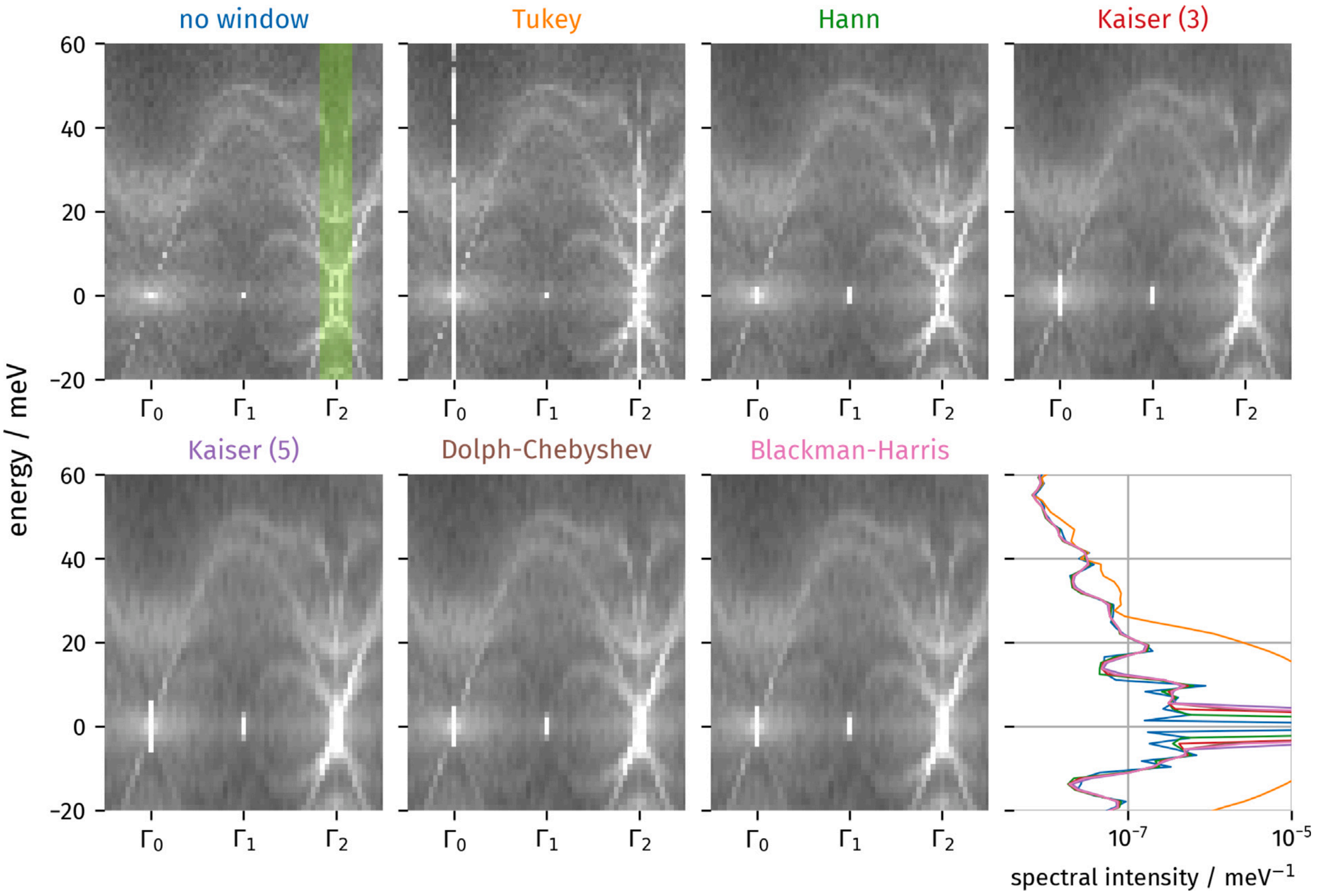

**Fig. A.7.** More examples of windowing functions.

In general, for timestep $\tau$, the obtainable energies will range from $-\epsilon$ to $\epsilon$, where

$$\epsilon = \frac{h}{2\,\tau} \quad \text{with a rule-of thumb} \quad \frac{\epsilon}{\text{meV}} \sim 2000 \left(\frac{\tau}{\text{fs}}\right)^{-1} \tag{B.1}$$

The timestep should be set so that all relevant energies are covered. It is recommended to cover an energy range at least 10 % to 20 % wider than the actual range of phonon energies to limit the effects of non-physical spectral wrap-around due to discrete sampling.

The number of timesteps in a trajectory chunk influences the energy resolution, since more timesteps result in a longer trajectory and consequently a finer frequency (energy) grid for a given time between samples. Increasing the number of samples per chunk decreases the influence of artifacts from windowing. On the other hand, the noise per frequency bin is increased as the statistical sample is smaller (within a trajectory with given length).

*Reciprocal space grids.* The shape of reciprocal (`kspace`) arrays for multislice is the same as the shape of the corresponding real-space simulation arrays for the wave function. It should be set so that the size of a pixel in real space is $\sim 0.05$ Å.

*Region of interest (ROI).* As it was discussed above, ROI slicing is critical for the code's performance. By narrowing down the chosen ROI in reciprocal space and in energy, the total required memory (RAM), disk space, and data flow can be significantly reduced. It is advised to choose ROIs only as big as really necessary.

*Optimal settings.* The speed of the code from the hardware perspective is mainly influenced by the available memory, either in the form of VRAM (GPU memory) or RAM (when running on CPUs). By changing the settings detailed in this section, we can heavily influence the memory requirements per computational batch. In order to optimize the calculation speed, we need to maximize the use of available memory.

In practice, there are two main parameters that can be changed to influence the total calculation time. The first of them is the trajectory-chunk length — lower values lead to lower memory requirements and, simultaneously, higher computing speed. However, by lowering it, we are decreasing the sampling in the energy axis.

The other parameter that one can change is the scanning batch shape. This parameter is purely computational and changing it does not influence the results in any way — it only influences how many scan points are processed in parallel. The `scanning_batch_shape` should thus be the parameter of choice when trying to optimize the speed of computation. I.e., after all other parameters are set to their desired values, we recommend to change the `scanning_batch_shape` so that the total memory required per batch is as large as it is allowed by the computing hardware.

The minimum memory requirements of holding all exit wave functions in (V)RAM can be estimated as follows. First we calculate the product of

$$s_x \cdot s_y \cdot k_x \cdot k_y \cdot c \cdot p \cdot 2, \tag{B.2}$$

where `scanning_batch_shape` $= [s_x, s_y]$; $[k_x, k_y]$ is the shape of kspace ROI (computed by the `Config` object when setting up the calculation, and stored in `simulation.kspace.ROI_shape` in config.yaml); $c$ is chunk length, $p$ is the precision, i.e., 16 B for double precision; and the trailing 2 is needed as torch lacks an in-place implementation of the FFT. In the case of the STEM-EELS calculation presented in Section 4, this amounts to $4 \times 8 \times 383 \times 383 \times 200 \times 16 \times 2\,\text{B} \approx 30\,\text{GB}$. But the exit wave functions are not the only data which needs to be held in memory. Significant space is also taken by the transmission functions, which are evaluated for each multislice run and whose memory requirements can be calculated in a similar manner. In this particular example, the transmission functions will require approximately 7 GB. One should also consider a reasonable overhead for other minor memory allocations.

Listing 1 : Example configuration script.

```
project_dir = '~/projects/torched-tacaw/'
moldyn_traj_file = project_dir + 'data/TiO2_20x20x80_stacked.traj'
data_dir = project_dir + 'data/tacaw/'
name = 'TiO2_20x20x80_stacked'
config_file = data_dir + 'config.yaml'

config = Config(
        name                        = name,
        datafolder                  = data_dir,
        config_file                 = config_file,

        beam_energy_keV             = 60.,
        beam_conv_ang_mrad          = 30,

        scanning_mode               = 'box',
        scanning_origin             = [ 0.0 , 0.0 ],
        scanning_box_end            = [ 4.6617899 , 4.6617899 ],
        scanning_shape              = [ 16, 16 ],
        scanning_batch_shape        = [ 4, 8 ],

        sample_temperature_K        = 300,

        trajectory_file             = moldyn_traj_file,
        trajectory_timestep_fs      = 15.0,
        trajectory_chunks_size      = 200,
        trajectory_chunks_skip_init = 200,

        kspace_shape_full           = [20*84,20*84],
        kspace_ROI_mode             = 'minmax_mrad',
        kspace_ROI_min              = [-100, -100],
        kspace_ROI_max              = [ 100,  100],

        frequency_THz_ROI           = [-30,30],

        n_slices                    = 2 * 80 ,
        intensities_zarray          = data_dir + 'tacaw.zarray',

        comments = [
             'TiO2␣structure␣20x20x80',
             'stacked␣from␣8␣20x20x10'
        ]

        dump_to_yaml = True,
    )

    config.create_zarr_array()
```

*Planewave calculation.* One can easily carry out planewave calculations with optimal k-space resolution, useful for phonon band structure measurements, by setting `convergence_angle_mrad` = 0 and `scanning_shape` = `scanning_batch_shape` = [0, 0].

### B.2. Running the calculation

When the configuration is set-up and the config-file saved, the calculation is ready to be executed. An example script that can be used for this purpose can look like

```
dispatcher = Dispatcher(
    config_file,
    device='cuda:0',
)
dispatcher.run()
```

Here, if the device is passed as an argument, it will overwrite the torch device from the config file. In principle one can run as many instances of this script as one wants to run in parallel and a practical limit is set only by the total number of computational batches and the disk write speed. The particular setup is heavily system-dependent. For example some of the calculations presented in Section 3 utilized the Dardel supercomputer hosted by Parallelldatorcentrum (PDC) at the KTH Royal Institute of Technology in Stockholm, and we typically run a similar script 8 times in parallel per GPU node with `device=f'cuda:{gpu_id}'`, where `gpu_id=0...7`, as each of Dardel's GPU nodes has 8 GPU cores. CPU devices can also be used.

For a detailed logging of the progress of the calculation, the parameter `logger` = `logging.Logger` can be passed to the `Dispatcher` object instantiation.

*STEM EELS detectors.* In vibrational STEM EELS experiments, one typically uses a specific aperture positioned in reciprocal plane to choose a solid angle within which the signal is collected. In `torched-TACAW` this functionality is implemented through the `DetectorSet` object. As its name suggests, it can be used to calculate STEM-EELS datasets for several detectors in parallel. It can be used for example in a script similar to this:

```
detector_set = DetectorSet(
    'config.yaml',
    dict(type = 'circular',
         center = [70,0],
         radius = 20,
         label = 'off-axis'),
    dict(type = 'circular',
         center = [0,0],
         radius = 20,
         label = 'BF'),
    dict(type = 'annular',
         center = [0,0],
         radius_inner = 80,
         radius_outer = 100,
         label = 'ADF'),
)
detector_set.work()
```

As a final note, we want to emphasize that the sample scripts presented above are the python part of setting up and running the calculation. The specific way of running these scripts (e.g., using queue systems like SLURM) is naturally system dependent.

## Data availability

The code of torched-TACAW is available on github referenced in the paper. Computational scripts are in examples.

## References

[1] O.L. Krivanek, T.C. Lovejoy, N. Dellby, T. Aoki, R. Carpenter, P. Rez, E. Soignard, J. Zhu, P.E. Batson, M.J. Lagos, R.F. Egerton, P.A. Crozier, Vibrational spectroscopy in the electron microscope, Nature 514 (7521) (2014) 209–212.

[2] J.A. Hachtel, J. Huang, I. Popovs, S. Jansone-Popova, J.K. Keum, J. Jakowski, T.C. Lovejoy, N. Dellby, O.L. Krivanek, J.C. Idrobo, Identification of site-specific isotopic labels by vibrational spectroscopy in the electron microscope, Science 363 (6426) (2019) 525–528.

[3] R. Senga, Y.-C. Lin, S. Morishita, R. Kato, T. Yamada, M. Hasegawa, K. Suenaga, Imaging of isotope diffusion using atomic-scale vibrational spectroscopy, Nature 603 (7899) (2022) 68–72.

[4] N. Li, R. Shi, Y. Li, R. Qi, F. Liu, X. Zhang, Z. Liu, Y. Li, X. Guo, K. Liu, Y. Jiang, X.-Z. Li, J. Chen, L. Liu, E.-G. Wang, P. Gao, Phonon transition across an isotopic interface, Nat. Commun. 14 (1) (2023) 2382.

[5] J.C. Idrobo, A.R. Lupini, T. Feng, R.R. Unocic, F.S. Walden, D.S. Gardiner, T.C. Lovejoy, N. Dellby, S.T. Pantelides, O.L. Krivanek, Temperature measurement by a nanoscale electron probe using energy gain and loss spectroscopy, Phys. Rev. Lett. 120 (2018) 095901, URL https://link.aps.org/doi/10.1103/PhysRevLett.120.095901.

[6] M.J. Lagos, P.E. Batson, Thermometry with subnanometer resolution in the electron microscope using the principle of detailed balancing, Nano Lett. 18 (7) (2018) 4556–4563.

[7] J. Kikkawa, K. Kimoto, Optical and acoustic phonon temperature measurements using electron nanoprobe and electron energy loss spectroscopy, Phys. Rev. B 106 (2022) 195431.

[8] F.S. Hage, G. Radtke, D.M. Kepaptsoglou, M. Lazzeri, Q.M. Ramasse, Single-atom vibrational spectroscopy in the scanning transmission electron microscope, Science 367 (6482) (2020) 1124–1127, URL https://science.sciencemag.org/content/367/6482/1124.

[9] X. Yan, C. Liu, C.A. Gadre, L. Gu, T. Aoki, T.C. Lovejoy, N. Dellby, O.L. Krivanek, D.G. Schlom, R. Wu, X. Pan, Single-defect phonons imaged by electron microscopy, Nature 589 (7840) (2021) 65–69, URL http://www.nature.com/articles/s41586-020-03049-y.

[10] B. Haas, T.M. Boland, C. Elsässer, A.K. Singh, K. March, J. Barthel, C.T. Koch, P. Rez, Atomic-resolution mapping of localized phonon modes at grain boundaries, Nano Lett. 23 (13) (2023) 5975–5980.

[11] F.S. Hage, D.M. Kepaptsoglou, Q.M. Ramasse, L.J. Allen, Phonon spectroscopy at atomic resolution, Phys. Rev. Lett. 122 (2019) 016103, URL https://link.aps.org/doi/10.1103/PhysRevLett.122.016103.

[12] K. Venkatraman, B.D.A. Levin, K. March, P. Rez, P.A. Crozier, Vibrational spectroscopy at atomic resolution with electron impact scattering, Nat. Phys. 15 (12) (2019) 1237–1241.

[13] E.R. Hoglund, H.A. Walker, K. Hussain, D.-L. Bao, H. Ni, A. Mamun, J. Baxter, J.D. Caldwell, A. Khan, S.T. Pantelides, P.E. Hopkins, J.A. Hachtel, Nonequivalent atomic vibrations at interfaces in a polar superlattice, Adv. Mater. 36 (33) (2024) 2402925, URL https://advanced.onlinelibrary.wiley.com/doi/abs/10.1002/adma.202402925.

[14] X. Yan, P.M. Zeiger, Y. Huang, H. Sun, J. Li, C.A. Gadre, H. Yang, R. He, T. Aoki, Z. Zhong, Y. Nie, R. Wu, J. Rusz, X. Pan, Atomic-scale imaging of frequency-dependent phonon anisotropy, Nature 645 (8082) (2025) 893–899.

[15] B. Haas, C.T. Koch, P. Rez, Perspective on atomic-resolution vibrational electron energy-loss spectroscopy, Appl. Phys. Lett. 125 (15) (2024) 150502.

[16] H. Yang, Y. Zhou, G. Miao, J. Rusz, X. Yan, F. Guzman, X. Xu, X. Xu, T. Aoki, P. Zeiger, X. Zhu, W. Wang, J. Guo, R. Wu, X. Pan, Phonon modes and electron–phonon coupling at the FeSe/$SrTiO_3$ interface, Nature 635 (8038) (2024) 332–336.

[17] D. Kepaptsoglou, J.Á. Castellanos-Reyes, A. Kerrigan, J.A. do Nascimento, P.M. Zeiger, K.E. hajraoui, J.C. Idrobo, B.G. Mendis, A. Bergman, V.K. Lazarov, J. Rusz, Q.M. Ramasse, Magnon spectroscopy in the electron microscope, Nature (2025).

[18] B.D. Forbes, L.J. Allen, Modeling energy-loss spectra due to phonon excitation, Phys. Rev. B 94 (1) (2016) 014110, URL https://link.aps.org/doi/10.1103/PhysRevB.94.014110.

[19] R.J. Nicholls, F.S. Hage, D.G. McCulloch, Q.M. Ramasse, K. Refson, J.R. Yates, Theory of momentum-resolved phonon spectroscopy in the electron microscope, Phys. Rev. B 99 (9) (2019) 094105, URL https://link.aps.org/doi/10.1103/PhysRevB.99.094105.

[20] R. Senga, K. Suenaga, P. Barone, S. Morishita, F. Mauri, T. Pichler, Position and momentum mapping of vibrations in graphene nanostructures, Nature 573 (2019) 247–250, URL https://www.nature.com/articles/s41586-019-1477-8.

[21] C. Dwyer, Prospects of spatial resolution in vibrational electron energy loss spectroscopy: Implications of dipolar scattering, Phys. Rev. B 96 (22) (2017) 224102, URL https://link.aps.org/doi/10.1103/PhysRevB.96.224102.

[22] P. Rez, A. Singh, Lattice resolution of vibrational modes in the electron microscope, Ultramicroscopy 220 (2021) 113162, URL http://www.sciencedirect.com/science/article/pii/S0304399120303053.

[23] P.M. Zeiger, J. Rusz, Efficient and Versatile Model for Vibrational STEM-EELS, Phys. Rev. Lett. 124 (2) (2020) 025501, URL https://link.aps.org/doi/10.1103/PhysRevLett.124.025501.

[24] P.M. Zeiger, J. Rusz, Simulations of spatially and angle-resolved vibrational electron energy loss spectroscopy for a system with a planar defect, Phys. Rev. B 104 (2021) 094103.

[25] J.Á. Castellanos-Reyes, P.M. Zeiger, J. Rusz, Dynamical theory of angle-resolved electron energy loss and gain spectroscopies of phonons and magnons in transmission electron microscopy including multiple scattering effects, Phys. Rev. Lett. 134 (2025) 036402, URL https://link.aps.org/doi/10.1103/PhysRevLett.134.036402.

[26] H.A. Walker, T.W. Pfeifer, P.M. Zeiger, J.A. Hachtel, S.T. Pantelides, E.R. Hoglund, Pyslice: Routine vibrational electron energy loss spectroscopy prediction with universal interatomic potentials, 2026, arXiv:2602.10064.

[27] B.D. Forbes, A.V. Martin, S.D. Findlay, A.J. D'Alfonso, L.J. Allen, Quantum mechanical model for phonon excitation in electron diffraction and imaging using a Born-Oppenheimer approximation, Phys. Rev. B 82 (10) (2010) 104103.

[28] N.R. Lugg, B.D. Forbes, S.D. Findlay, L.J. Allen, Atomic resolution imaging using electron energy-loss phonon spectroscopy, Phys. Rev. B 91 (14) (2015) 144108, URL https://journals.aps.org/prb/abstract/10.1103/PhysRevB.91.144108.

[29] J.E. Lennard-Jones, Cohesion, Proc. Phys. Soc. 43 (5) (1931) 461.

[30] M.H. Müser, S.V. Sukhomlinov, L. Pastewka, Interatomic potentials: achievements and challenges, Adv. Phys.: X 8 (1) (2023) 2093129.

[31] S. Baroni, S. de Gironcoli, A. Dal Corso, P. Giannozzi, Phonons and related crystal properties from density-functional perturbation theory, Rev. Modern Phys. 73 (2001) 515–562.

[32] L. Zhang, Y. Wan, Y. Shibuta, X. Huang, Progress in machine learning interatomic potential and its applications in materials science, Prog. Nat. Sci.: Mater. Int. 35 (6) (2025) 1079–1104, URL https://www.sciencedirect.com/science/article/pii/S1002007125001492.

[33] D. Marx, J. Hutter, Ab Initio Molecular Dynamics: Basic Theory and Advanced Methods, Cambridge University Press, 2009.

[34] L. Barroso-Luque, S. Muhammed, X. Fu, B.M. Wood, M. Dzamba, M. Gao, A. Rizvi, C.L. Zitnick, Z.W. Ulissi, Open materials 2024 (omat24) inorganic materials dataset and models, 2024, arXiv preprint arXiv:2410.12771.

[35] J. Schmidt, N. Hoffmann, H.-C. Wang, P. Borlido, P.J. Carriço, T.F. Cerqueira, S. Botti, M.A. Marques, Machine-learning-assisted determination of the global zero-temperature phase diagram of materials, Adv. Mater. 35 (22) (2023) 2210788, URL https://onlinelibrary.wiley.com/doi/full/10.1002/adma.202210788.

[36] C. Chen, S.P. Ong, A universal graph deep learning interatomic potential for the periodic table, Nat. Comput. Sci. 2 (11) (2022) 718–728.

[37] M. Neumann, J. Gin, B. Rhodes, S. Bennett, Z. Li, H. Choubisa, A. Hussey, J. Godwin, Orb: A fast, scalable neural network potential, 2024, arXiv:2410.22570.

[38] I. Batatia, P. Benner, Y. Chiang, A.M. Elena, D.P. Kovács, J. Riebesell, X.R. Advincula, M. Asta, M. Avaylon, W.J. Baldwin, F. Berger, N. Bernstein, A. Bhowmik, F. Bigi, S.M. Blau, V. Cărare, M. Ceriotti, S. Chong, J.P. Darby, S. De, F.D. Pia, V.L. Deringer, R. Elijošius, Z. El-Machachi, F. Falcioni, E. Fako, A.C. Ferrari, J.L.A. Gardner, M.J. Gawkowski, A. Genreith-Schriever, J. George, R.E.A. Goodall, J. Grandel, C.P. Grey, P. Grigorev, S. Han, W. Handley, H.H. Heenen, K. Hermansson, C. Holm, C.H. Ho, S. Hofmann, J. Jaafar, K.S. Jakob, H. Jung, V. Kapil, A.D. Kaplan, N. Karimitari, J.R. Kermode, P. Kourtis, N. Kroupa, J. Kullgren, M.C. Kuner, D. Kuryla, G. Liepuoniute, C. Lin, J.T. Margraf, I.-B. Magdău, A. Michaelides, J.H. Moore, A.A. Naik, S.P. Niblett, S.W. Norwood, N. O'Neill, C. Ortner, K.A. Persson, K. Reuter, A.S. Rosen, L.A.M. Rosset, L.L. Schaaf, C. Schran, B.X. Shi, E. Sivonxay, T.K. Stenczel, V. Svahn, C. Sutton, T.D. Swinburne, J. Tilly, C. van der Oord, S. Vargas, E. Varga-Umbrich, T. Vegge, M. Vondrák, Y. Wang, W.C. Witt, T. Wolf, F. Zills, G. Csányi, A foundation model for atomistic materials chemistry, 2025, arXiv:2401.00096.

[39] B. Rhodes, S. Vandenhaute, V. Šimkus, J. Gin, J. Godwin, T. Duignan, M. Neumann, Orb-v3: atomistic simulation at scale, 2025, arXiv:2504.06231.

[40] B.M. Wood, M. Dzamba, X. Fu, M. Gao, M. Shuaibi, L. Barroso-Luque, K. Abdelmaqsoud, V. Gharakhanyan, J.R. Kitchin, D.S. Levine, K. Michel, A. Sriram, T. Cohen, A. Das, A. Rizvi, S.J. Sahoo, Z.W. Ulissi, C.L. Zitnick, UMA: A family of universal models for atoms, 2026, arXiv:2506.23971.

[41] J. Riebesell, R.E.A. Goodall, P. Benner, Y. Chiang, B. Deng, G. Ceder, M. Asta, A.A. Lee, A. Jain, K.A. Persson, A framework to evaluate machine learning crystal stability predictions, Nat. Mach. Intell. 7 (6) (2025) 836–847.

[42] M. Osmera, J. Rusz, torched-TACAW, 2026, Available at https://github.com/Martexiwell/torched-tacaw. archived at https://doi.org/10.5281/zenodo.20929266.

[43] F. Bigi, M.F. Langer, M. Ceriotti, The dark side of the forces: assessing non-conservative force models for atomistic machine learning, in: Proceedings of the 42nd International Conference on Machine Learning, in: Proceedings of Machine Learning Research, 267, PMLR, 2025, pp. 4384–4414, URL https://proceedings.mlr.press/v267/bigi25a.html.

[44] F. Heinz, S. Jähnigen, J.-L. Schäfer, B.G. Keller, How thermostats influence dynamics across time scales: A systematic study from fast motions to slow transitions, 2025, arXiv:2511.22631.

[45] G. Bussi, T. Zykova-Timan, M. Parrinello, Isothermal-isobaric molecular dynamics using stochastic velocity rescaling, J. Chem. Phys. 130 (7) (2009) 074101.

[46] P.K. Patra, B. Bhattacharya, Nonergodicity of the nose-hoover chain thermostat in computationally achievable time, Phys. Rev. E 90 (2014) 043304.

[47] T.W. Ko, J.A. Finkler, S. Goedecker, J. Behler, A fourth-generation high-dimensional neural network potential with accurate electrostatics including non-local charge transfer, Nat. Commun. 12 (1) (2021) 398.

[48] S. Grimme, J. Antony, S. Ehrlich, H. Krieg, A consistent and accurate ab initio parametrization of density functional dispersion correction (DFT-d) for the 94 elements H-pu, J. Chem. Phys. 132 (15) (2010) 154104.

[49] J. Barthel, Dr. Probe: A software for high-resolution STEM image simulation, Ultramicroscopy 193 (2018) 1–11, URL https://www.sciencedirect.com/science/article/pii/S0304399118301402.

[50] H. Brown, Py_multislice, 2026, http://dx.doi.org/10.5281/zenodo.5762735, Available at: https://github.com/HamishGBrown/py_multislice (Last Accessed 12 February 2026).

[51] J. Madsen, T. Susi, The abTEM code: transmission electron microscopy from first principles, Open Res. Eur. 1 (2021) 24.

[52] A. Paszke, S. Gross, F. Massa, A. Lerer, J. Bradbury, G. Chanan, T. Killeen, Z. Lin, N. Gimelshein, L. Antiga, A. Desmaison, A. Kopf, E. Yang, Z. DeVito, M. Raison, A. Tejani, S. Chilamkurthy, B. Steiner, L. Fang, J. Bai, S. Chintala, Pytorch: An imperative style, high-performance deep learning library, in: Advances in Neural Information Processing Systems 32, Curran Associates, Inc., 2019, pp. 8024–8035, URL http://papers.neurips.cc/paper/9015-pytorch-an-imperative-style-high-performance-deep-learning-library.pdf.

[53] A. Hjorth Larsen, J. Jørgen Mortensen, J. Blomqvist, I.E. Castelli, R. Christensen, M. Duł ak, J. Friis, M.N. Groves, B. Hammer, C. Hargus, E.D. Hermes, P.C. Jennings, P. Bjerre Jensen, J. Kermode, J.R. Kitchin, E. Leonhard Kolsbjerg, J. Kubal, K. Kaasbjerg, S. Lysgaard, J. Bergmann Maronsson, T. Maxson, T. Olsen, L. Pastewka, A. Peterson, C. Rostgaard, J. Schiøtz, O. Schütt, M. Strange, K.S. Thygesen, T. Vegge, L. Vilhelmsen, M. Walter, Z. Zeng, K.W. Jacobsen, The atomic simulation environment—a python library for working with atoms, J. Phys.: Condens. Matter. 29 (27) (2017) 273002.

[54] Zarr, Zarr (version 3), 2026, Available at: https://zarr.readthedocs.io (Last Accessed 18 February 2026).

[55] W. Van den Broek, X. Jiang, C. Koch, FDES, a GPU-based multislice algorithm with increased efficiency of the computation of the projected potential, Ultramicroscopy 158 (2015) 89–97.

[56] I. Lobato, D. Van Dyck, An accurate parameterization for scattering factors, electron densities and electrostatic potentials for neutral atoms that obey all physical constraints, Acta Crystallogr. Sect. A Found. Adv. 70 (6) (2014) 636–649.

[57] O. Ben-Kiki, C. Evans, I. döt Net, YAML ain't markup language (YAML) version 1.2.2, 2021, YAML specification. https://yaml.org/spec/1.2.2/.

[58] P. Welch, The use of fast Fourier transform for the estimation of power spectra: A method based on time averaging over short, modified periodograms, IEEE Trans. Audio Electroacoust. 15 (2) (1967) 70–73.

[59] G. Bussi, D. Donadio, M. Parrinello, Canonical sampling through velocity rescaling, J. Chem. Phys. 126 (1) (2007) 014101.

[60] Z. He, J. Rusz, Temperature-dependent vibrational EELS simulations with nuclear quantum effects, 2026, arXiv:2603.20744.

[61] J.Á. Castellanos-Reyes, I.P. Miranda, P.M. Zeiger, A. Bergman, J. Rusz, Theory of momentum-resolved electron energy-loss spectra of coupled phonon and magnon excitations, 2025, arXiv:2508.07073.

[62] P. Virtanen, R. Gommers, T.E. Oliphant, M. Haberland, T. Reddy, D. Cournapeau, E. Burovski, P. Peterson, W. Weckesser, J. Bright, S.J. van der Walt, M. Brett, J. Wilson, K.J. Millman, N. Mayorov, A.R.J. Nelson, E. Jones, R. Kern, E. Larson, C.J. Carey, İ. Polat, Y. Feng, E.W. Moore, J. VanderPlas, D. Laxalde, J. Perktold, R. Cimrman, I. Henriksen, E.A. Quintero, C.R. Harris, A.M. Archibald, A.H. Ribeiro, F. Pedregosa, P. van Mulbregt, SciPy 1.0 Contributors, SciPy 1.0: Fundamental Algorithms for Scientific Computing in Python, Nature Methods 17 (2020) 261–272.